\documentclass[a4paper,11pt]{article}
\usepackage{pos}
\usepackage{subfigure}
\usepackage{braket}
\title{Accelerated Quantum Circuit Monte-Carlo Simulation for Heavy Quark Thermalization}

\author*[a]{Wenyang Qian}

\affiliation[a]{Instituto Galego de F\'isica de Altas Enerx\'ias IGFAE, \\
Universidade de Santiago de Compostela,
E-15782 Galicia-Spain}

\emailAdd{qian.wenyang@usc.es}

\abstract{Heavy quark thermalization in the quark-gluon plasma (QGP) is one of the most promising phenomena for understanding the strong interaction, where their energy loss and momentum broadening at low momentum can be well described by a stochastic process with drag and diffusion terms. We propose an accelerated quantum circuit Monte-Carlo (aQCMC) framework that ultilizes the quantum amplitude estimation (QAE) algorithm to simulate heavy quark thermalization with quadratically less resources. Specifically, we simulate the thermalization of a heavy quark in both 1D and 2D and in isotropic and anisotropic mediums using an ideal quantum simulator and compare that to analytical thermal expectations.}

\FullConference{12th Large Hadron Collider Physics Conference  (LHCP2024)\\
3-7 June 2024 \\
Boston, USA \\}

\begin{document}
\maketitle

\section{Introduction}

Quantum computing (QC) technology has already been extensively applied in the high-energy physics (HEP) community~\cite{ Bauer:2022hpo}. Novel gate-based quantum finance strategy~\cite{Woerner_2019} based on the \emph{quantum amplitude estimation} (QAE)~\cite{brassard2002quantum} shows a promising quadratic speed-up over the classical Monte-Carlo (MC) method. Our work is the first QC application using QAE to study heavy-quark thermalization. Due to their heavy masses, the heavy quarks thermalize, in the QGP background, can be well-described by a stochastic process with low-momentum random kicks from the medium~\cite{Du:2022uvj}. In this description, the thermalization is controlled by two competing effects, the energy loss from a drag term and a diffusion from a stochastic term. In a non-relativistic or static limit, it follows to a classical Maxwell-Boltzmann distribution. For our simulations, various events are simulated as quantum state evolution and physical observables are extracted with amplitude estimation.

\section{Theoretical setup}

In this work, the heavy-quark thermalization is characterized by the stochastic differential equation (SDE) known as the Langevin equation. In position-momentum phase space $\vec{x},\vec{p}$,
\begin{align}
dx_i&=\frac{p_i}{E(\vec{p})}dt,\quad dp_i=-A(\vec{x},\vec{p},t)p_{i}dt+\sigma_{ij}(\vec{x},\vec{p},t)dW_j,\quad i=x,y,z,
\label{eq-langevin}
\end{align}
where $t$ is the evolution time, $dW\sim\mathcal{N}(0,dt)$ is the Wiener process describing the random sampling that satisifies the correlation $\braket{dW_idW_j}=\delta_{ij}dt$, $A(\vec{x},\vec{p},t)$ is the drag coefficient, and $\sigma_{ij}(\vec{x},\vec{p},t)$ is the diffusion coefficient. These coefficients may be calculated from experiments and phenomelogies. In the non-relativistic limit with a heavy quark mass $M$ at a small momentum $p$, the Langevin equation is simplified and can be written in a dimensionless way:
\begin{align}
dq_i=-q_id\tilde{t}+d\tilde{W}_i.
\label{eq:langevin_unity}
\end{align}
where $q_i=p_i/M$, $d\tilde{t}=Adt=\sigma_{ii}^2\chi_{i}^2/(2MT)dt$, $d\tilde{W}_i\sim\mathcal{N}(0,2Td\tilde{t}/(M\chi_i^2))$ are dimensionless variables. Here, we use $d\tilde{t}\simeq 1/N_t$ so it takes about $N_t$ steps for the thermalization. The thermal distribution at the equilibrium in turn becomes $f^{\rm eq}(\vec{q})\propto\exp(-q_x^2/\tilde{\sigma}_x^2-q_y^2/\tilde{\sigma}_y^2-q_z^2/\tilde{\sigma}_z^2)$.

The stochastic process of heavy quark thermalization is usually computed with the Monte-Carlo (MC) methods, by sampling the Wiener process for each time step $\tilde{t}$ for $\vec{q}^{\tilde{t}}$ over a collection of events $N_\mathrm{event}$ to produce an emergent phenomenon of heavy quark thermalization. The physical observable $F$ is extracted by $\braket{F}=\sum_{i=1}^{N_{\rm event}} F(\vec{q}^{\tilde{t}})_{\mathrm{event}=i}/N_{\rm event}.$

\section{Quantum strategy to speedup thermalization}

\begin{figure}
    \centering
    \subfigure[\;The aQCMC circuit\label{fig:circuit}]{
    \includegraphics[width=0.47\textwidth]{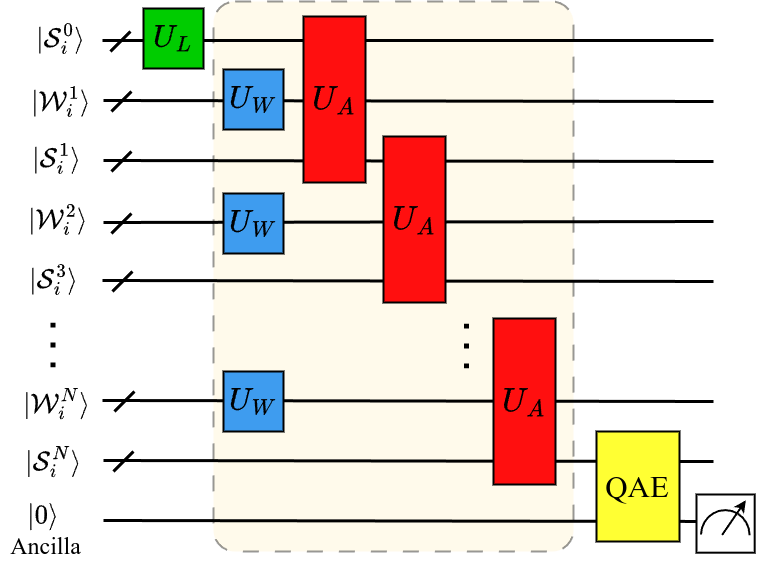}
    }
    \subfigure[\;Quantum speedup\label{fig:speedup}]{
    \includegraphics[width=0.47\textwidth]{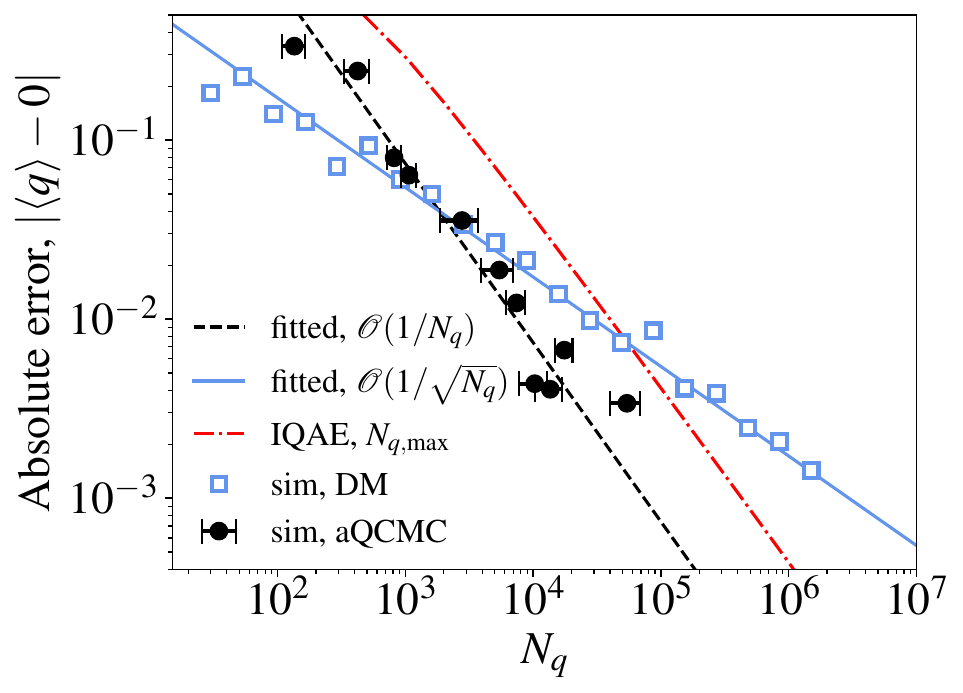}
    }
    \caption{The aQCMC algorithm: (a) schematic quantum circuit and (b) demonstration of quantum speedup.
    \label{fig:aQCMC}}
\end{figure}

\begin{table*}[t]
\centering
\begin{tabular}{|c|c|c|c|c|c|c|c|c|}
\hline
Index & 0 & 1 & 2 & 3 & $\cdots$  & $2^n-2$ & $2^n-1$ \\ \hline
$\ket{\psi}$ & $\ket{0...000}$ & $\ket{0...001}$ & $\ket{0...010}$ & $\ket{0...011}$ & $\cdots$  & $\ket{1...110}$ & $\ket{1...111}$ \\ \hline
$q$ & $-q_{\rm max}$ & $-q_{\rm max}+\delta q$ & $-q_{\rm max}+2\delta q$ & $q_{\rm max}+3\delta q$ & $\cdots$  & $q_{\rm max}-2\delta q$ & $q_{\rm max}-\delta q$ \\ \hline
$\bar {q}$ & $0$ & $\delta q$ & $2\delta q$ & $3\delta q$ & $\cdots$ & $2q_{\rm max}-2\delta q$ & $2q_{\rm max}-\delta q$ \\ \hline
\end{tabular}
\caption{\label{tab:encoding} Binary qubit encoding of the momenta $q$ on a $n$-qubit register in each spatial dimension.
}
\end{table*}
We propose a new quantum strategy, i.e., the \emph{accelerate quantum circuit Monte-Carlo} (aQCMC) method, to simulate the heavy quark thermalization in a stochastic description. 
In aQCMC, we encode the particle's momenta $q_i$ in each spatial direction $i = x,y,z$ as a quantum state $\ket{\psi}$ (see Table.~\ref{tab:encoding}). Using $n$ qubits, we restrict $q\in[-q_{\rm max},q_{\rm max})$ and discretize $q$ into $2^n$ values with $\delta q=2q_{\rm max}/2^n$. In addition, we shift $q$ to the positive $\bar{q}$ by a constant $q_\mathrm{max}$ and impose a periodic boundary condition. We use quantum register $\mathcal{S}_i^t$ to encode $q_i$ and $\mathcal{W}^t_i$ to encode $d\tilde{W}_i$ at each time step $t$ for each direction $i$ (see Fig.~\ref{fig:circuit}). The \emph{distribution loading gate ($U_L$)} loads the initial momentum distribution $\ket{q^0_i}$. The \emph{stochastic Wiener gates ($U_W$)} provide the stochastic normal distribution $\mathcal{N}(0,\sigma=2Td\tilde{t}/(M\chi_i^2))$ for each time step. The \emph{quantum evolution gates ($U_{A^n_i}$)} at step $n$ calculates the drag and diffusion terms in Eq.~\eqref{eq:langevin_unity}. Lastly, the QAE\cite{brassard2002quantum} block evaluates $\braket{F}$ of the final state $\ket{q_i^N}_{\mathcal{S}^{N}_i}$ on the ancilla using $N_q$ queries at an error $\epsilon = \mathcal{O}(1/N_q)$, a quadratic improvement over classical MC, and it is convincingly demonstrated in Fig.~\ref{fig:speedup}.

\section{Results}

\begin{figure}
    \centering
    \subfigure[\label{fig:QAE_p} Expectation $\braket{q}$]{
    \includegraphics[width=0.23\textwidth]{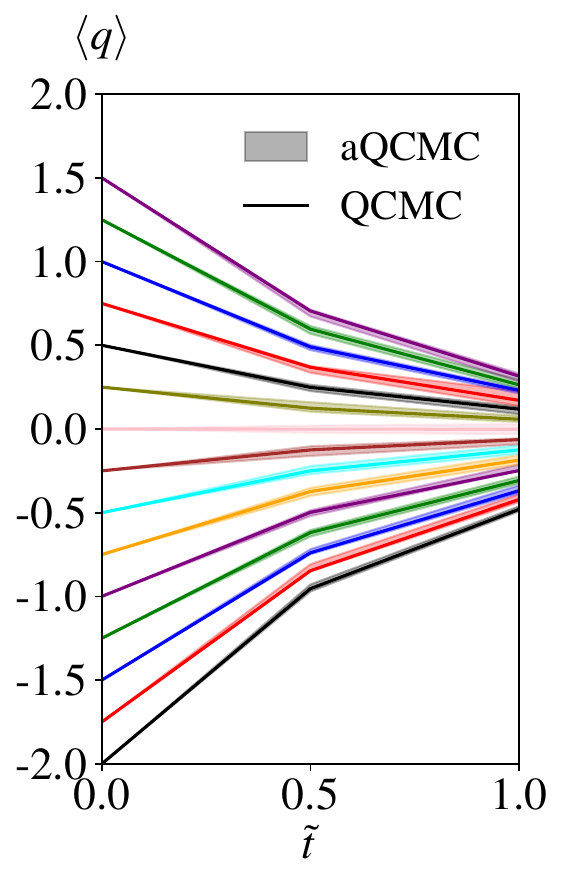}}
    \subfigure[\label{fig:QAE_Absp} Expectation $\braket{|q|}$]{
    \includegraphics[width=0.23\textwidth]{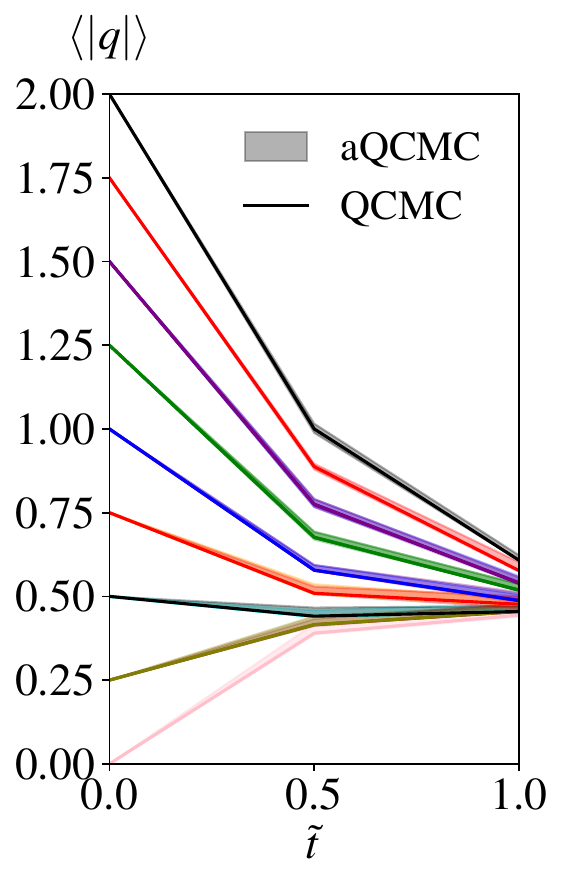}}
    \subfigure[\label{fig:qc_v2} Elliptic flow $v_2$]{
    \includegraphics[width=0.40\textwidth]{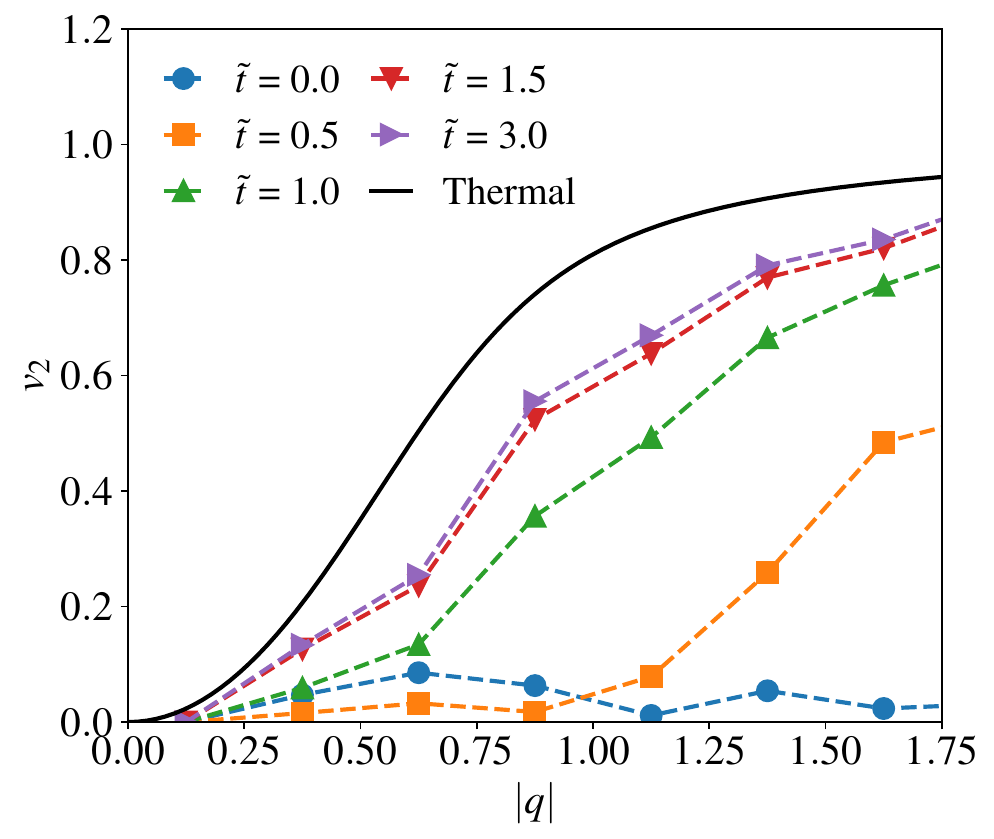}}
    
    \caption{\label{fig:result}
    Quantum simulation using the aQCMC with QAE for (a,b) 1D and (c) 2D heavy quark thermalizations. The aQCMC results agree with direct measurement (QCMC) for earlier time steps.
    }
\end{figure}

We present the numerical simulation results of the 1D and 2D heavy quark thermalization using the $\mathsf{QASM}$ simulator provided by $\mathsf{Qiskit}$. Specically, the heavy quark mass $M=1.5$\,GeV in a typical plasma temperature of $T\simeq 300$ - $500$\,MeV in HIC with $\tilde{\sigma}_i^2d\tilde{t}=2Td\tilde{t} / (M\chi_i^2)\simeq 2d\tilde{t}/ (5\chi_i^2)$ - $2d\tilde{t}/({3\chi_i^2})$. In Fig.~\ref{fig:result}, we present the simulation results using registers of sizes $|{\mathcal{S}}| = |{\mathcal{W}}| = 4$ qubits, so $q\in [-q_{\mathrm{max}}, -q_{\mathrm{max}}) = [-2, 2)$ with $\delta q=0.25$ and $d\tilde{t} = 0.5$. We simulate for $N_{\tilde{t}}=3$ steps, taking up a total of 20 qubits (and 1 ancilla for QAE). In Figs.~\ref{fig:QAE_p} and \ref{fig:QAE_Absp}, we observe the thermalization of the heavy quark's momentum and absolute momentum as an attractor towards thermal equilibirum. Extending to 2D systems with anisotropic mediums of $\chi_x=1$, $\chi_y=2$ (or $\tilde{\sigma}^2_{xx}=1/2$, $\tilde{\sigma}^2_{yy}=1/8$) we observe in Fig.~\ref{fig:qc_v2} the build up of the elliptic flow $v_2$ that characterizes the anisotropization, which agrees with analytical thermal equilibirum at late time.

\section{Conclusion and Outlook}
In this work, we propose the aQCMC strategy to accelerate the computation for heavy quark thermalization in both 1D and 2D mediums. We show their thermalization patterns and late-time behaviors are comparable to the analytical expectations, using quadratically fewer resources. It remains interesting to extend this to quarkonium dissociation and recombination in the medium.

\section*{Acknowledgement}
We are grateful to X. Du who has made important contributions to this work~\cite{Du:2023ewh}. WQ is supported by European Research Council under ERC-2018-ADG-835105 YoctoLHC; by Maria de Maeztu excellence unit grant CEX2023-001318-M and project PID2020-119632GB-I00 funded by MICIU/AEI/10.13039/501100011033; by ERDF/EU; and by MSCA under Grant 101109293.

\bibliographystyle{plain}
\bibliography{mybib.bib}

\begin{thebibliography}{1}

\bibitem{Bauer:2022hpo}
Christian~W. Bauer et~al.
\newblock {Quantum Simulation for High-Energy Physics}.
\newblock {\em PRX Quantum}, 4(2):027001, 2023.

\bibitem{brassard2002quantum}
Gilles Brassard, Peter Hoyer, Michele Mosca, and Alain Tapp.
\newblock Quantum amplitude amplification and estimation.
\newblock {\em Contemporary Mathematics}, 305:53--74, 2002.

\bibitem{Du:2023ewh}
Xiaojian Du and Wenyang Qian.
\newblock {Accelerated quantum circuit Monte~Carlo simulation for heavy quark thermalization}.
\newblock {\em Phys. Rev. D}, 109(7):076025, 2024.

\bibitem{Du:2022uvj}
Xiaojian Du and Ralf Rapp.
\newblock {Non-equilibrium charmonium regeneration in strongly coupled quark-gluon plasma}.
\newblock {\em Phys. Lett. B}, 834:137414, 2022.

\bibitem{Woerner_2019}
Stefan Woerner and Daniel~J. Egger.
\newblock Quantum risk analysis.
\newblock {\em npj Quantum Information}, 5(1), feb 2019.

\end{thebibliography}



\end{document}